\documentclass[twocolumn, twocolappendix]{aastex631}

\usepackage{graphicx}	
\usepackage{amsmath}	
\usepackage{bm}
\usepackage{newtxtext,newtxmath}
\usepackage[T1]{fontenc}
\usepackage{comment}
\usepackage[normalem]{ulem}
\usepackage{CJK}
\DeclareRobustCommand{\VAN}[3]{#2}
\let\VANthebibliography\thebibliography
\def\thebibliography{\DeclareRobustCommand{\VAN}[3]{##3}\VANthebibliography}
\usepackage{subfigure}

\renewcommand{\epsilon}{\varepsilon}

\renewcommand{\epsilon}{\varepsilon}

\renewcommand{\epsilon}{\varepsilon}

\def\rg{r_{\rm g}}

\begin{document}

\title{Once a giant, (almost) always a giant: Partial Tidal Disruption Events of Giant Stars}
\author{N\'uria Navarro Navarro} 
\author[0000-0002-7964-5420]{Tsvi Piran}
\affiliation{Racah Institute for Physics, The Hebrew University, Jerusalem, 91904, Israel}

\correspondingauthor{N\'uria Navarro Navarro\ ; \ Tsvi Piran}
\email{nuria.navarronav@mail.huji.ac.il ; tsvi.piran@mail.huji.ac.il}

\begin{abstract}
Tidal disruption events (TDEs) of  giant stars by supermassive black holes (SMBH) differ significantly from those of main sequence ones.   Most (all for SMBH of more than $ {\rm a~ few }\times 10^5 m_\odot$) giant-TDEs are partial:  only a fraction of the envelope is torn apart. The dense stellar core and the rest of the envelope remain intact.  In this work we explore, using the stellar evolution code {\texttt {MESA}}, the fate of the remnants. We find that after a short period, comparable to the thermal time scale, the remnant returns to a giant structure with a radius comparable to the progenitor giant one, a slightly larger luminosity (as compared with a regular giant with the same mass) and a comparable lifetime until it collapses to a white dwarf.  If such a giant with mass less than $\approx 0.9 m_\odot$ is discovered  it can be identified as an outlier - a giant that is too light for the current age of the Universe.  If the remnant orbit is not perturbed significantly during the encounter, the remnant will undergo successive partial tidal disruptions until its mass is $0.6-0.7 m_\odot$. We expect a few dozen to a few hundred such remnants in the Galactic nucleus. 
\end{abstract}


\section{Introduction}
\label{sec:intro}

A Tidal Disruption Event (TDE)  occurs when a star that has wandered into the vicinity of a supermassive black hole (SMBH) gets torn apart by the SMBH's tidal force. As a result, a fraction of the mass falls onto the black hole, leading to a luminous flare, while the remaining debris becomes unbound and escapes to infinity. 

Even though giants compose a smaller fraction of stars, giant-TDEs will be quite frequent. Because of their 
{ much lower average density giants will be disrupted at much larger distances than main sequence (MS) star. } The probability of disruption is proportional to the tidal radius, which in turn is proportional to the stellar radius. Thus,  a significant fraction of TDEs involve the disruption of a giant.  Still, the disruption of giant stars has been little explored \citep[see however][]{macleodTidalDisruptionGiant2012,bogdanovicDISRUPTIONREDGIANT2014,Guillochon2014,rossiProcessStellarTidal2021}.

The tidal disruption of a giant star is very different from the well-studied disruption of a MS star. A giant has a very dense core surrounded by a low-density envelope.   The sharp difference in density between these two regions implies that the outer layers of the envelope can easily be peeled off while the compact core can remain effectively unperturbed. This results in a partial tidal disruption (PTDE) in which only a fraction of the envelope is torn apart and the core and the inner envelope remain as a surviving remnant. 
Because of the large density of giants' cores a total disruption is unlikely unless the SMBH is rather light, $M_{_{\rm BH}} \lesssim 3 \times 10^5 m_\odot$.
 
A natural question arises: what is the long-term evolution of the surviving remnant? Can we spot an observational signature that will enable us to identify it as a surviving  TDE victim? 
This question is also of interest in the context of globular clusters, where the alleged presence of intermediate black holes (IMBH) would lead  to PTDEs of giant stars  and, therefore, to remnant victim stars. Such remnants may indicate the presence of an intermediate black hole at the cluster's core.

With this motivation, we study the long-term evolution of the surviving  victims of giant PTDEs, exploring a range of different star masses,  stellar ages, and stripped mass fractions. We focus on giants in the horizontal branch  that  burn He in their core. For more massive stars $M> 1.5 m_\odot$  it is most likely to catch a giant at this phase which is the longest in its giant evolution. The subgiant phase is longer for less massive stars, but the stellar size is typically much smaller then and the chances for a TDE are smaller. 
Giants are larger and are more susceptible to disruption during their AGB and red giant phases, however, these phases are much shorter so it is less likely to have a TDE  then. 
The structure of this work is as follows:  we introduce the basics of tidal disruption events, reviewing some well-understood order-of-magnitude derivations and the partial tidal disruption of giants in  \S \ref{sec:tdes}. In  \S \ref{sec:methods} we describe the methodology, approximations and simulations done using {\texttt {MESA}}. In  \S \ref{sec:results} we present and discuss our results. We conclude by summarizing our findings in \S \ref{sec:conclusions}.

\section{Full and Partial Tidal disruptions}
\label{sec:tdes}

A star that passes near a SMBH is torn apart 
if the tidal forces of the SMBH are larger than the star's self-gravity \citep{Hills1975,reesTidalDisruptionStars1988}. The approximate distance from the SMBH at which this happens is the tidal radius:
\begin{equation} 
R_{\rm t }\simeq R_\star  \left( \frac{M_{_{\rm BH}}} {M_\star} \right)^{1/3} \ , 
\label{eq:rt} 
\end{equation}
where $M_{_{\rm BH}}$ is the SMBH mass and $M_\star$, $R_\star$ are the stellar mass and radius, respectively.  The duration, $\approx \left({R_{\rm t }^3}/{GM_{_{\rm BH}}}\right)^{1/2}$,  that the star spends near $R_t$,   is comparable to the dynamical time of the star $t_{dyn}\approx (G\bar \rho_\star)^{-1/2}$ (where $\bar \rho_\star$ is the average stellar density)\footnote{Note that \citet{ryuTidalDisruptionsMainsequence2020} have shown that for realistic MS  stars, Eq. ~\ref{eq:rt} provides a good estimate of the maximal tidal radius for a full disruption. They also obtained corrections factors of order unity for this equation. Clearly, these corrections are not relevant for 
{ giant stars} and we won't use them. }. 

\begin{figure}
\includegraphics[width=.47\textwidth]{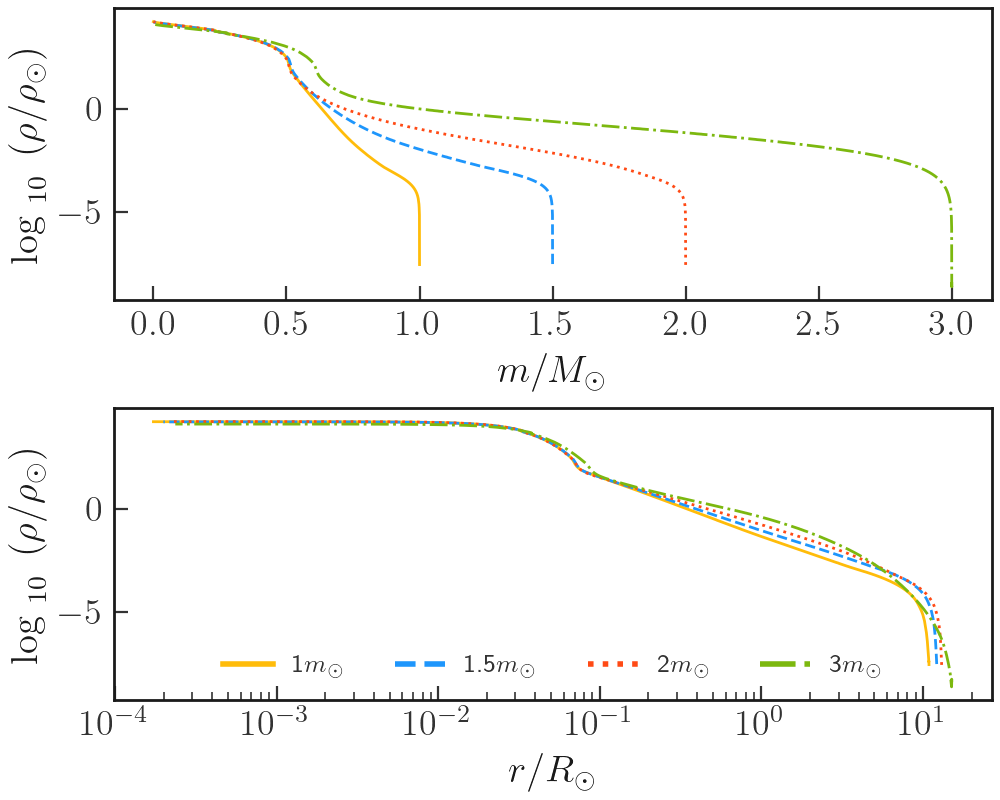}
    \centering
    \caption{{ Density profiles as a function of mass and radius for $1m_\odot$, $1.5m_\odot$, $2m_\odot$ and $3m_\odot$ {giants in the horizontal branch} at the time when the core He fraction $f_{_{\rm He}}=0.50$, compared to solar parameters (mass, $m_\odot$, radius, $R_\odot$ and average density, $\rho_\odot$). During this stage, the core is $\sim10^6$ times denser than the envelope.}}
    \label{fig:density_ms_rg}
\end{figure}

Giants have a very dense core and a tenuous envelope (see Fig. ~\ref{fig:density_ms_rg}). While for MS stars the ratio of core density to average density is of the order $\rho^{\text{MS}}_c/\bar{\rho}^{\text{MS}}_\star\approx 10^2$, for giants the ratio is four orders of magnitude greater $\rho^\text{ giant}_c/\bar{\rho}^{\text{giant}}_\star\approx10^6$. Therefore, the estimate given by Eq.~\ref{eq:rt} is invalid
for  TDEs of giants. Specifically, the core for low and intermediate mass stars (1-3$m_\odot$) has a mass $M_{\rm c} \approx 0.5 m_\odot$ and a radius of about $10^9$ cm.  The tidal distance needed to disrupt the core is  $\approx 100$ times smaller than the estimate given by Eq.~\ref{eq:rt}.  {The giant core is very compact and comparable to a WD and it cannot be disrupted by SMBHs more massive than} $\approx 3 \times 10^5 M_\sun$ \citep[see e.g.][]{Krolik2011}. 
For $M_{_{\rm BH}}\gtrsim 3 \times 10^5 M_\sun$ all giant TDEs are partial \citep{macleodTidalDisruptionGiant2012,rossiProcessStellarTidal2021},  only  part of the envelope is torn apart.

\begin{figure}
\includegraphics[width=0.47\textwidth]{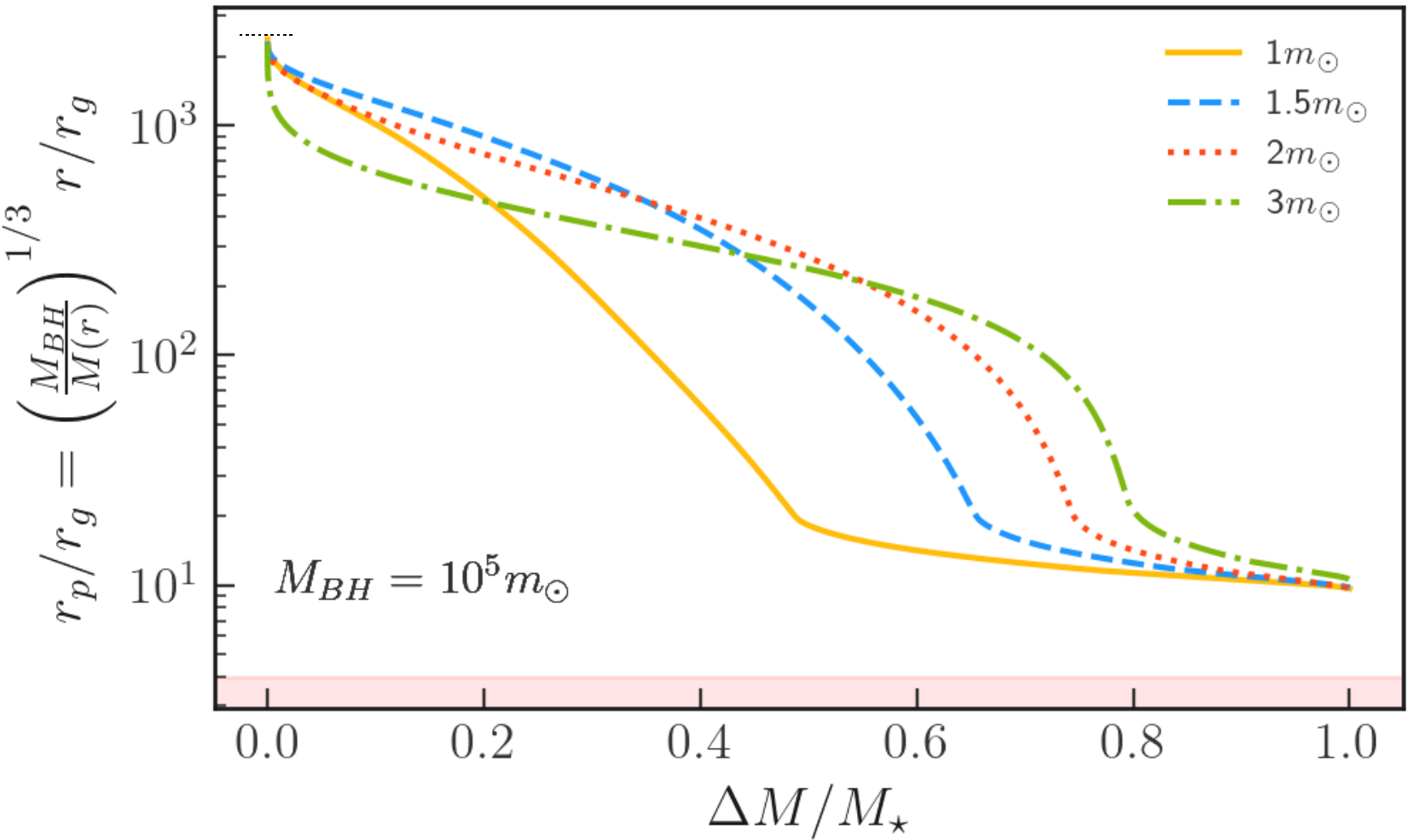}

\includegraphics[width=0.47\textwidth]{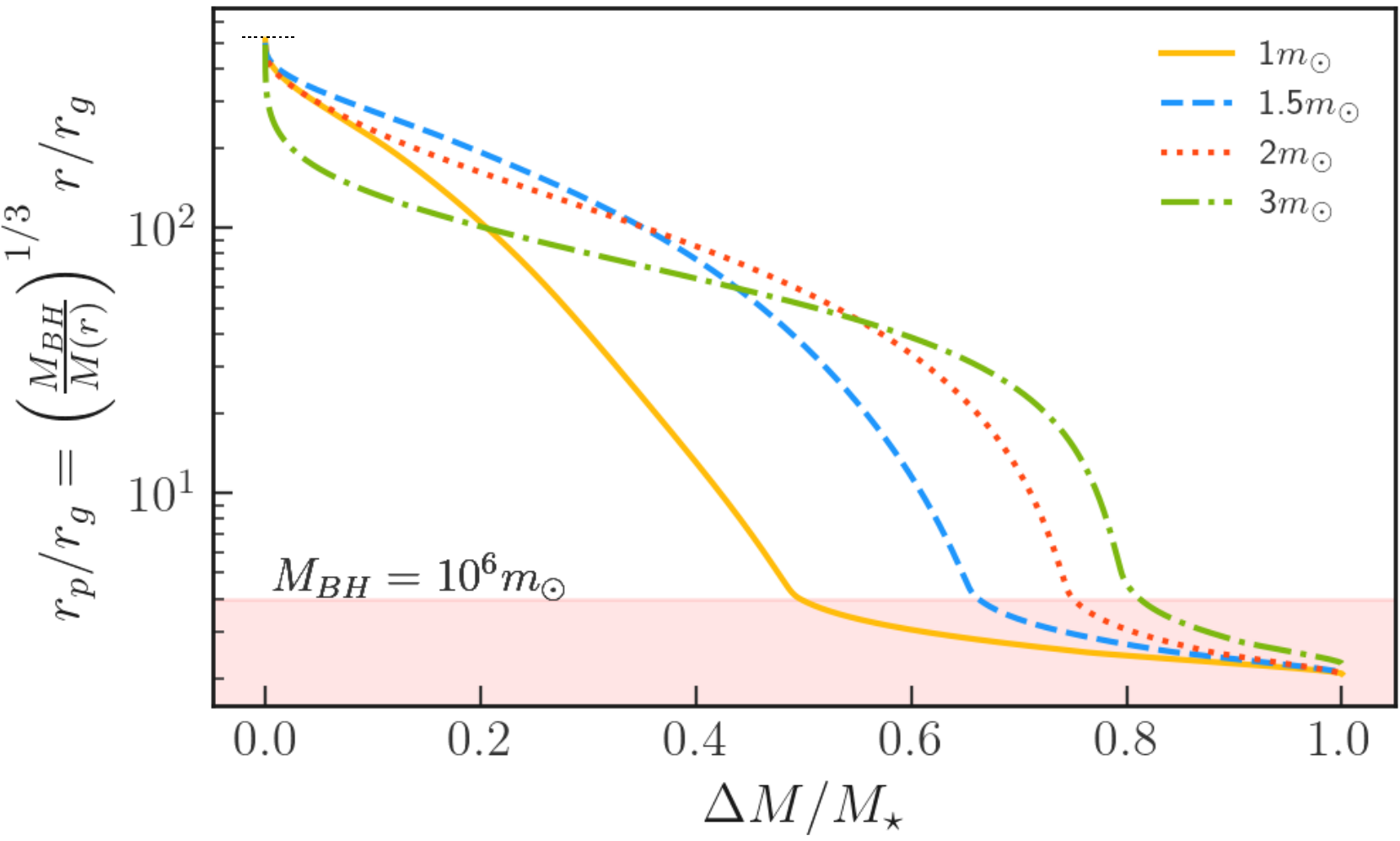}

\includegraphics[width=0.47\textwidth]{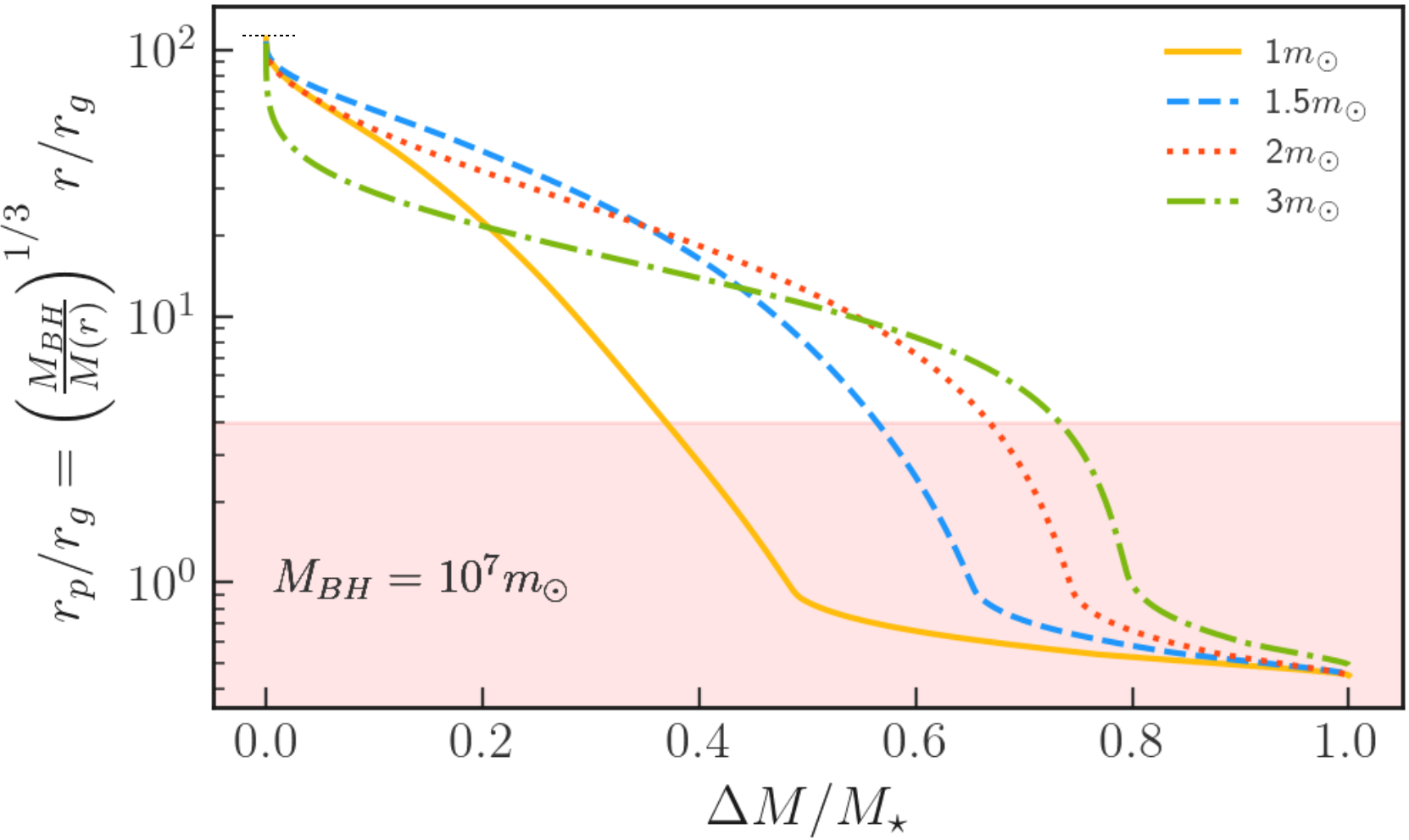}
    \caption{The orbital pericenter distance, $r_{\rm p}$, in units of the gravitational radius $\rg$ as a function of the stripped mass for  horizontal branch giants with $f_{_{\rm He}}=0.50$ and  different masses  and for different SMBH masses. The red region marks $4 \rg$ which is the minimal pericenter distance for an orbit with zero orbital energy around a Schwarzchild black hole. For a comparision $R_{\rm t}$ given by Eq. 1 is marked by a thin black dashed dotted. }
    \label{fig:td_bhmass}
\end{figure}

To estimate the mass that will be torn apart in  a partial disruption event, we use the orbit's pericenter, $r_{\rm p}$.  The mass that remains bound  is given by the approximate relation \citep{ryuTidalDisruptionsMainsequence2020a}: 
\begin{equation}
M(r) = \frac{4 \pi}{3} \int_0^r \rho(r') r'^2 dr' \approx \left( \frac{r}{r_{\rm p}}\right)^3 M_{_{\rm BH}} \ , 
\label{eq:mr}
\end{equation}
where  $M(r)$ is the giant mass up to radius r \footnote{{ \citep{ryuTidalDisruptionsMainsequence2020a} also calculate a correction factor, of order unity, for this formula. However, their correction factor is relevant only for MS stars.}}.   The ejected mass satisfies   $\Delta M=M_\star-M(r)$. 

\begin{figure}   \includegraphics[width=.45\textwidth]{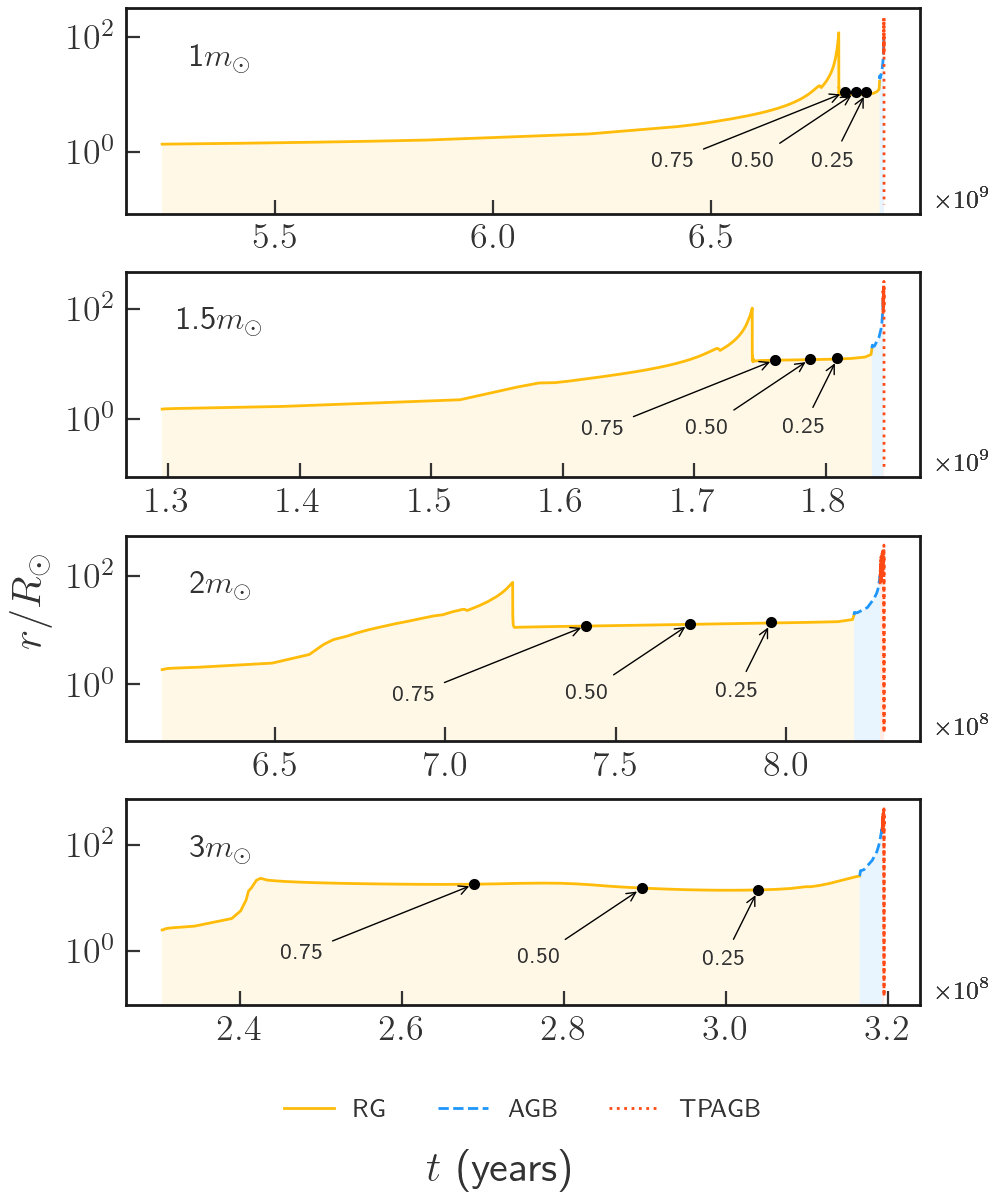}
    \centering
    \vspace{0.1in}
    \caption{Radius as a function of time during the late phases of the evolution (TAMS to WD) of the stars that we study, $1m_\odot$, $1.5m_\odot$, $2m_\odot$ and $3m_\odot$ with metallicity $0.1Z_\odot$. The black dots indicate the three stripping times that we will consider later. Those times are characterized by the He fraction ($f_{\rm He}=$ 0.75, 0.5 and 0.25).}
    \label{fig:radius_time_all}
\end{figure} 

Fig.~\ref{fig:td_bhmass} 
depicts the pericenter distance, $r_{\rm p}$, in gravitational radius units, $r_{\rm g}={GM_{_{\rm BH}}}/{c^2}$, as a function of the stripped mass fraction $\Delta M/{M_\star}$, using realistic horizontal branch giant profiles calculated with the {\texttt {MESA} stellar evolution code (shown in Fig.~\ref{fig:density_ms_rg}). The red-shaded area marks the minimal pericenter for parabolic orbits around a Schwarzschild black hole, $4r_{\rm g}$. Inside this region, the star's core would plunge into the black hole. Note, however, that when the core plunges into the black hole, some of the envelope will be torn apart before the star approaches the SMBH, and this material will not plunge directly into the SMBH. The implications of this will be explored elsewhere. 
For a SMBH with mass $M_{_{\rm BH}}=10^6m_\odot$ to strip half of the mass of a $1m_\odot$ star, the star would have to reach the minimal pericenter distance of $4 \rg$. A larger mass fraction, up to $\approx 0.8$ for $3m_\odot$, can be stripped from more massive stars.

{An important corollary of Eq.~\ref{eq:mr} and Fig.~\ref{fig:td_bhmass} is that for a significant mass to be torn apart from a giant in PTDE, $r_{\rm p }$, the pericenter of the giant's orbit should be significantly (at least a factor of 2) smaller than $R_{\rm t}$ calculated using Eq.~\ref{eq:rt}. As the resulting change in energy of the torn apart material is proportional\footnote{The change in energy is proportional to $r/r^2_{\rm p}$, where $r$ that is given by Eq.~\ref{eq:mr} is $ \propto r_{\rm p}$.} to  $R_{\rm t} /r_{\rm p}$   estimates of the energy of the unbound material or of the return time of the bound material should be revised accordingly. }

\section{Methods}
\label{sec:methods}
Our goal is to determine the evolution of the  remnants following a partial TDE of a giant star  \citep[see e.g.][for a dicussion of the different stages of  evolution of RG stars]{Kippenhahn1990} .  To do so, we strip a fraction of the mass of {giants in the horizontal branch} and evolve the stripped remnant. 
This is done in three stages. 

In the first stage we evolve, using the open source code  Modules for Experiments in Stellar Astrophysics  \citep[{\texttt {MESA}}, version r23.05.1;][see the Appendix for the implementation of {\texttt {MESA}} including details about convection, winds, and the rest of parameters]{Paxton2011, Paxton2013, Paxton2015, Paxton2018, Paxton2019, Jermyn2023},  the stars from the pre-main sequence stage up until they become white dwarfs.  
The internal structure of different stars is shown in Fig.~\ref{fig:density_ms_rg}. Fig.
\ref{fig:radius_time_all} describes the evolution of the radius vs time of the unperturbed stars. It also
depicts the stellar sizes at the moments used for stripping.  We have chosen  giants in the horizontal branch because most of the giant phase is in this phase for stars with mass $M>1.5m_\odot$ and it is more likely to have TDEs at this stage for less massive stars, as mentioned in the introduction. For example, even though a $1 m_\odot$, $0.1Z_\odot$ star reaches a radius $R_{\star} \sim 80R_\odot$ during the red giant phase, this stage constitutes only about $\sim 10^6$ years out of $\sim 1.5\times 10^9$ years of its life as a giant.

\begin{figure}  
\centering
\includegraphics[width=0.45\textwidth]{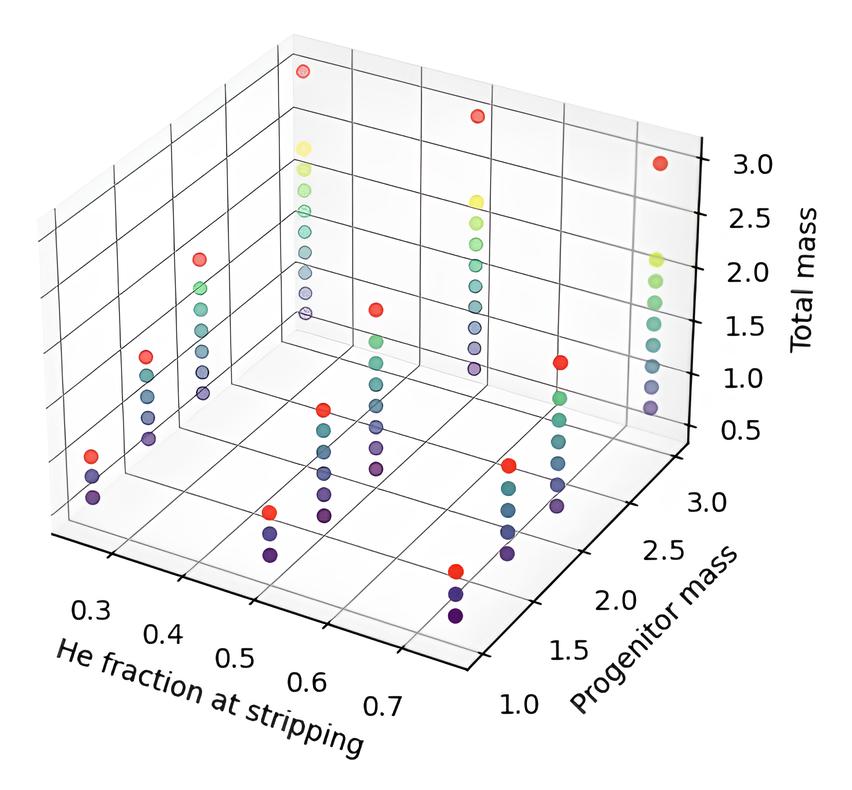}
    \caption{Each dot corresponds to a configuration simulated. { The configurations are characterized by the progenitor's mass, the total mass after disruption and the He fraction at stripping. } Red dots { describe unstripped }stars, the rest are stars that have been stripped.}
    \label{fig:parspace}
\end{figure}

In the second stage, we stop the regular evolution. We use the Helium fraction in the core, $f_{_{\rm He}}=0.75,0.5$ and $0.25$, as a stopping condition. Recall that as He is burning at the core this fraction decreases with time (see Fig.~\ref{fig:radius_time_all}).  We then artificially introduce a wind that removes a part of the envelope, keeping spherical symmetry, using the {\texttt {relax\_mass\_to\_remove\_H\_env}} routine in {\texttt{MESA}}. 
Indeed, this stage should be done using a full 
3D hydro simulations. This would have been essential if the envelope had a chemical gradient, since  a partial disruption event can mix chemical elements. However, because giant envelopes consist of hydrogen, the hydrodynamical effects of disruption can be ignored and the boosted mass loss via wind would make essentially the same outcome.
As we are interested in the fate of the self-bound remnant, the assumption of spherical symmetry is reasonable.


\begin{figure}  \includegraphics[width=0.45\textwidth]{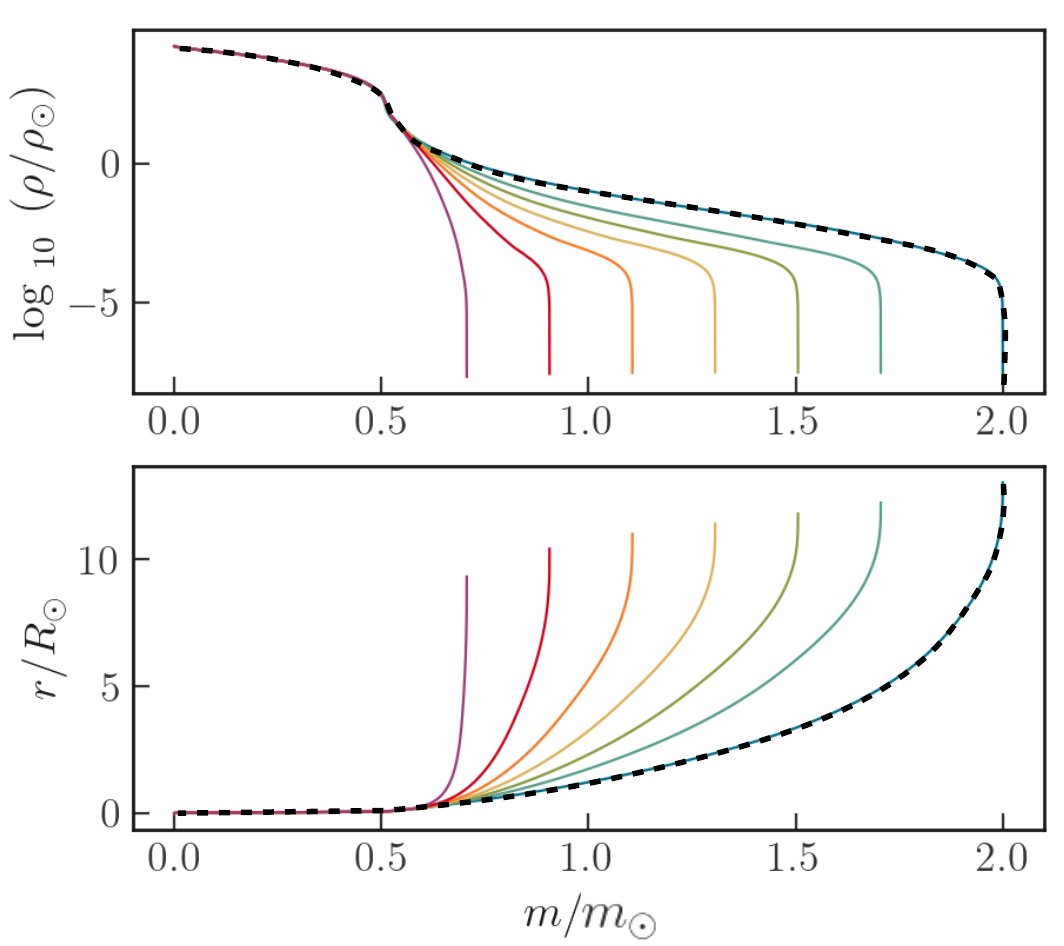}
    \caption{Density and radius vs. the Lagrangian mass coordinate before (dashed-black line) and after stripping and  thermal relaxation of a $2m_\odot$ star, for different stripped masses $\Delta M = 0, 0.3, 0.5,  0.7, 0.9, 1.1, 1.3 m_\odot$ or $\Delta M/M =0, 0.15, 0.25,  0.35, 0.45, 0.55, 0.65 $ (from right to left).} 
    \label{fig:2m_profile}
\end{figure}

Hydrodynamic and thermal equilibrium of the remnant are maintained in this process which takes place over a thermal time scale of the giant star. The former is reasonable since the dynamical time of the remnant is shorter than or comparable to $t_{\rm p}= r_{\rm p}^{3/2}/(G M_{_{\rm BH}})^{1/2}$,  the pericenter passage time. The latter is unrealistic since the pericenter passage time is much shorter than the thermal timescale, $t_p \ll t_{_{KH}} \sim 10^5$ yrs. Thus,  our simulations describe the remnant stars {\it after reaching thermal equilibrium}, that is, after $\sim 10^5$ years.
Comparisons of the stars before and after this stripping process are shown in Fig.~\ref{fig:2m_profile}.

\begin{figure}
\centering
\includegraphics[width=0.5\textwidth]{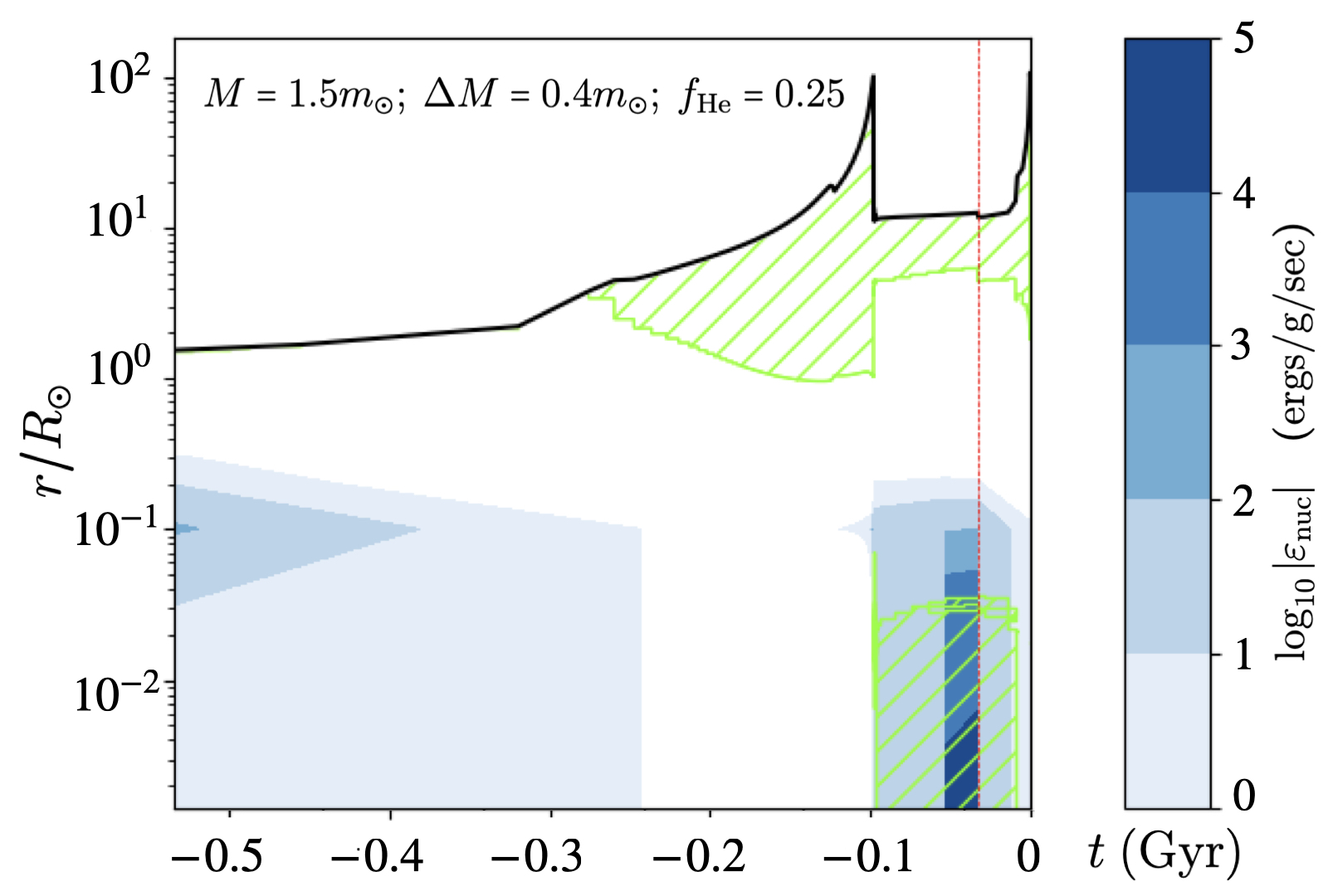}
\hfill\vspace{0.1in}
\vskip-0.4cm
\includegraphics[width=0.5\textwidth]{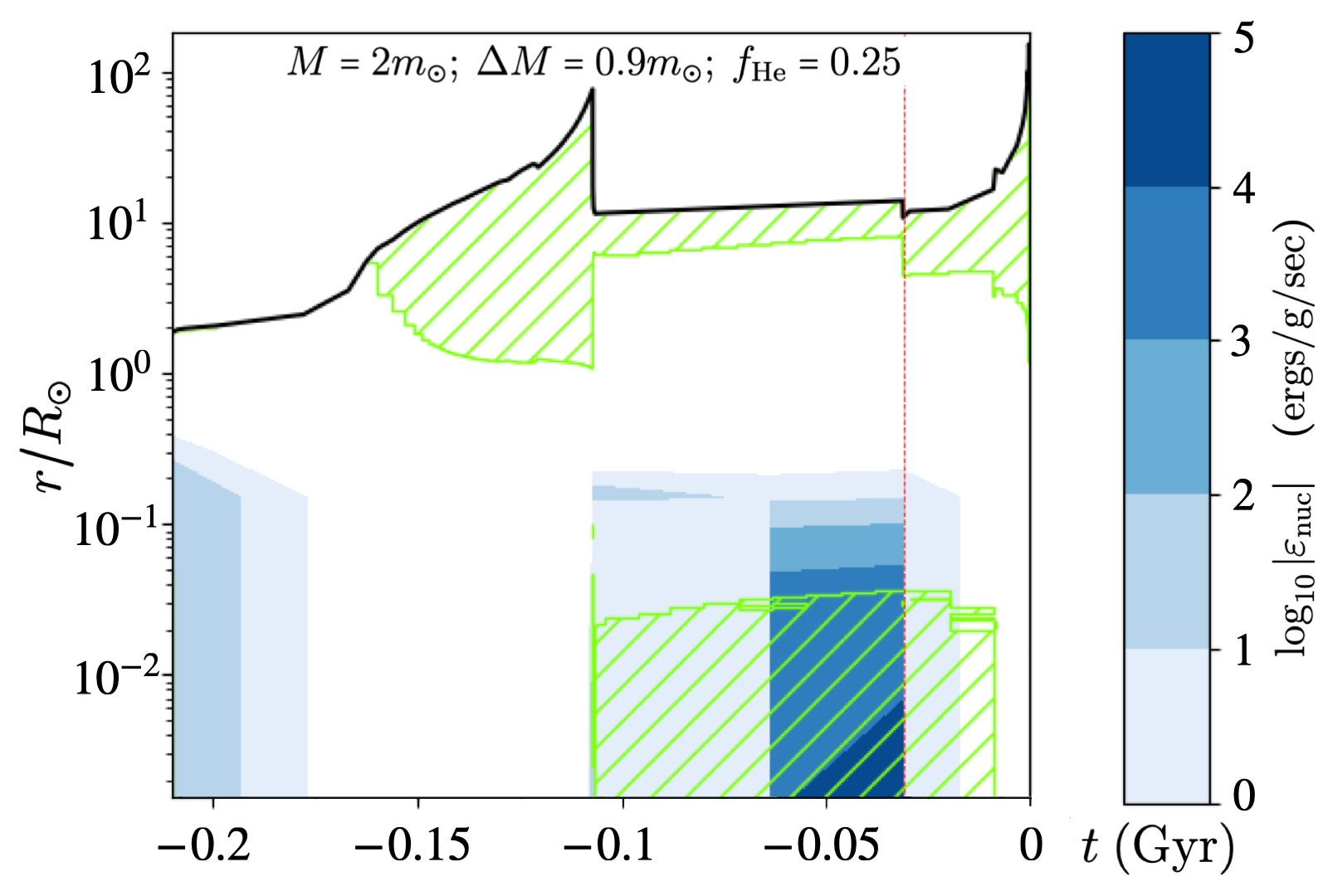}
\hfill\vspace{0.1in}
\vskip-0.4cm
\includegraphics[width=0.5\textwidth]{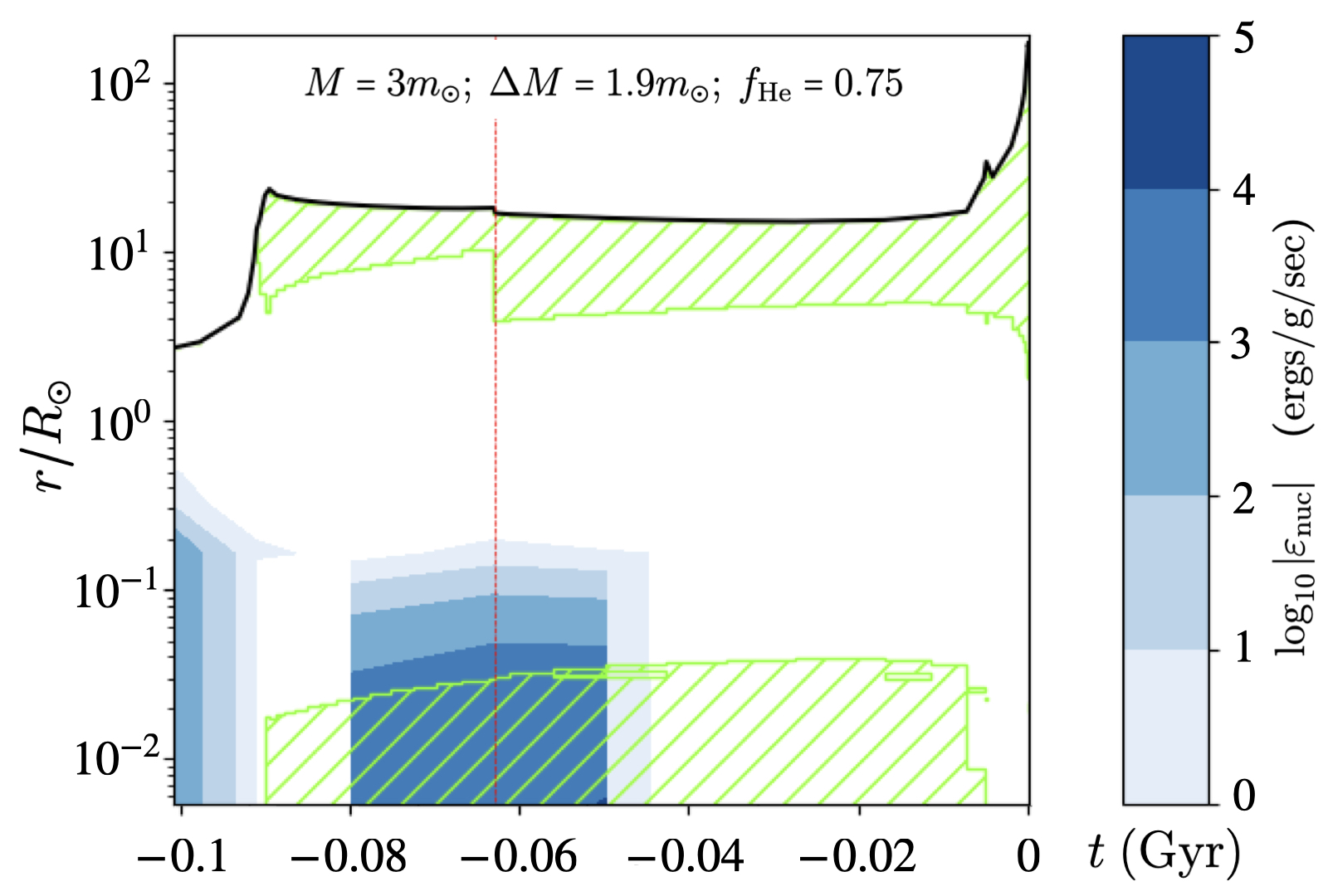}
    \caption{Kippenhahn diagrams from the beginning of the red giant phase until the stars enter the WD phase of representative stripped stars, with the same resulting mass of $1.1 m_\odot$. A dashed vertical red line marks the time of the stripping. Hatched green areas correspond to convection, and the blue shading describes the nuclear burning rate. Note the similarity in structure after stripping for the three cases. Time is shifted so that the collapse to a WD takes place at t=0.}
    \label{fig:kipp_stripped}
\end{figure}

In the third stage, once the chosen mass $\Delta M$ has been removed, we use  {\texttt {MESA}} to follow the subsequent evolution of the remnant until it becomes a white dwarf (WD). The evolution at this stage is the focus of this project.

\begin{figure*}
\includegraphics[width=0.9\textwidth]{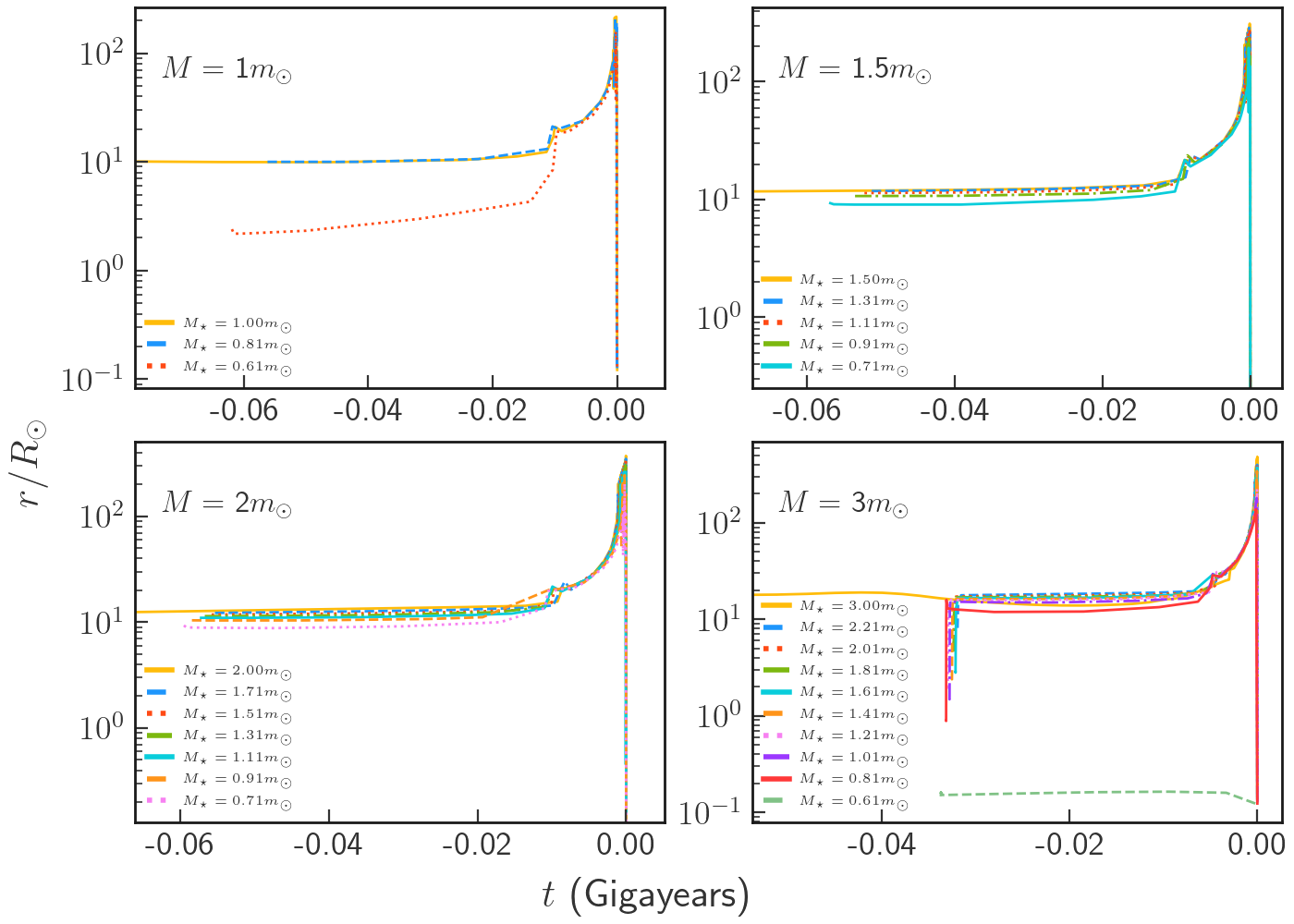}
    \centering
    \caption{Radius vs. time of stars stripped at $f_{_{\rm He}} =0.50$  compared to their progenitor for different progenitor masses and different total mass of the remnants, $M_*$. The time coordinate has been shifted so that all stars become WDs at the same time, which is set to $t=0$.}
    \label{fig:rad_time_050}
\end{figure*}

\begin{figure}
\includegraphics[width=0.45\textwidth]{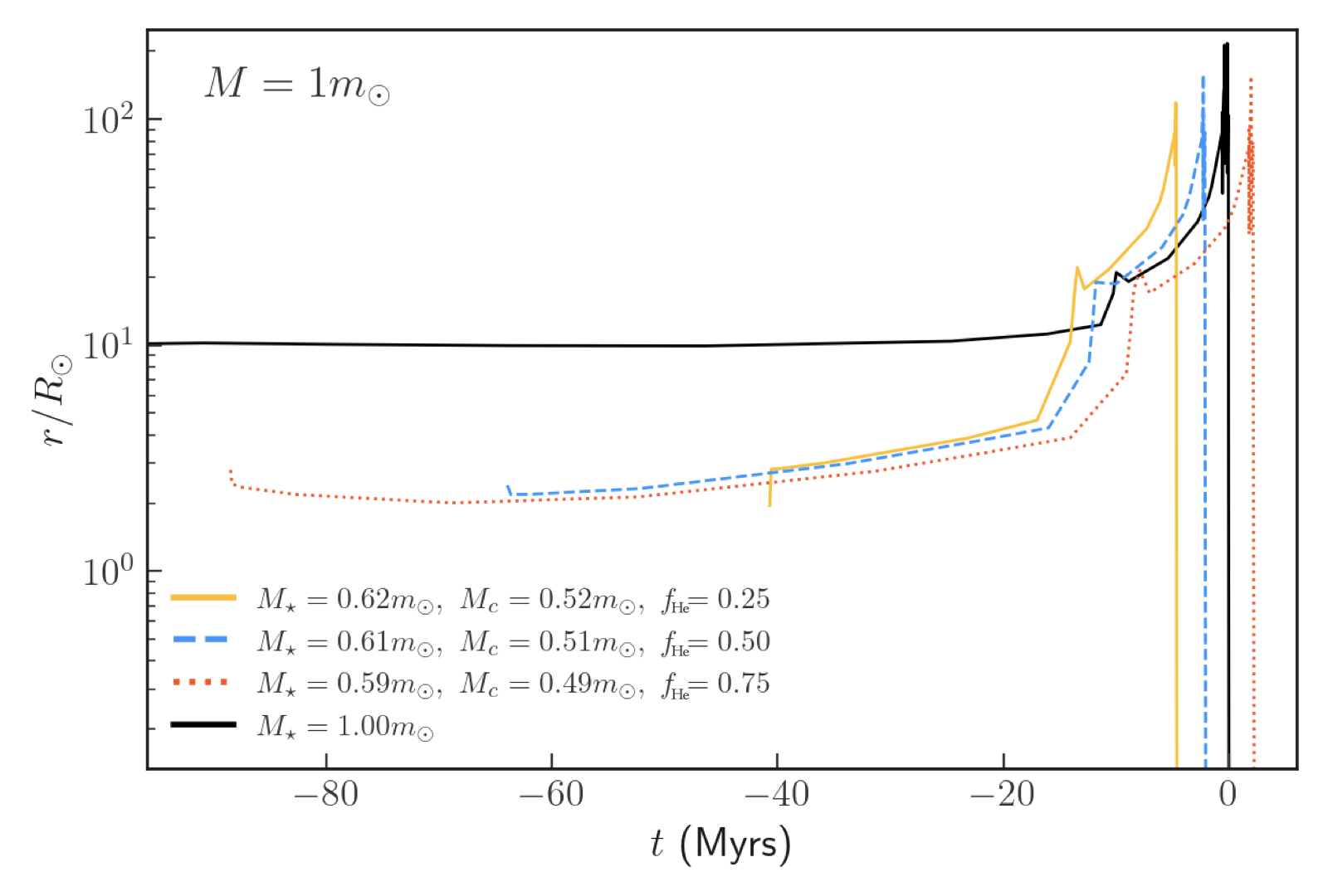}
    \centering
    \caption{Radius vs. time of stars stripped to the same $M_*$ at different $f_{_{\rm He}}$ compared to their 1 solar mass progenitor. 
    The time coordinate has been shifted so that the progenitor becomes WDs at $t=0$. Star that are stripped earlier (high $f_{\rm He}$ die later). }
    \label{fig:new}
\end{figure}

We have performed a total of 75 simulations, taking stars with a pre-main sequence mass of $1m_\odot$, $1.5m_\odot$, $2m_\odot$ and $3 m_\odot$ and metallicity\footnote{As we are interested in low mass stars we consider low metalicity as only such stars would have reached a giant phase at this age of the Universe.} of $Z=0.1Z_\odot$.  We have stripped different mass fractions at $f_{_{\rm He}}= 0.25$, $0.50$, and $0.75$ and evolved the stripped stars until they reach the WD phase. 
Fig.~\ref{fig:parspace} depicts the simulated parameter space.

\section{Results}
\label{sec:results}

The density profiles of the giant stars at $f_{_{\rm He}}= 0.5$  are shown in Fig. ~\ref{fig:density_ms_rg} for  giants just before stripping. The profiles at other stripping moments are rather similar. All the stars have a core of $0.5-0.6 m_\odot$ with the same extent of $\sim 0.04 R_\odot$. The envelope varies in density depending on the total mass, but the overall size is comparable $\sim 10 R_\odot$.

A comparison between the stars before and after stripping is shown in Fig.~\ref{fig:2m_profile}. One can see that unless  $\lesssim 0.15 m_\odot$ of the envelope mass is retained, the stripped stars recover a giant structure. The remaining envelope expands quickly and just becomes more tenuous. The outer layers reach more or less the same radius as before. 

Stars whose envelope has been completely stripped die immediately; that is, the bare core becomes a helium white dwarf in a brief period.  This case is dynamically unlikely as only SMBHs smaller than $\approx 3 \times 10^5 m_\odot$ can strip giants to that level, and even in those cases a very small $r_{\rm p}$ is required. In the following, we won't discuss these cases.  

Stars that partially retain their envelope continue their life {in the horizontal branch.} 
Fig.~\ref{fig:kipp_stripped} depicts the Kippenhahn diagrams of three different stripped stars that have the same mass after stripping. One can see that the stripped stars' internal structure is rather similar to the one of regular giants.  After the stripping, the core remains convective and burning  and the envelope is extended.  
The luminosity of the stripped stars is lower than the luminosity of their progenitor star (before stripping) but it is larger by a factor of $\sim 2$
than the luminosity of a giant with the same total mass. 

Fig. ~\ref{fig:rad_time_050} depicts the behavior of the stellar radius during the last stages. The time coordinate has been shifted so that the stars become white dwarfs at the same time. This allows us to see that the radius of the stripped stars is very similar to the progenitor's radius during the late stages of their evolution. 
\begin{figure*}
    \includegraphics[width=0.9\textwidth]{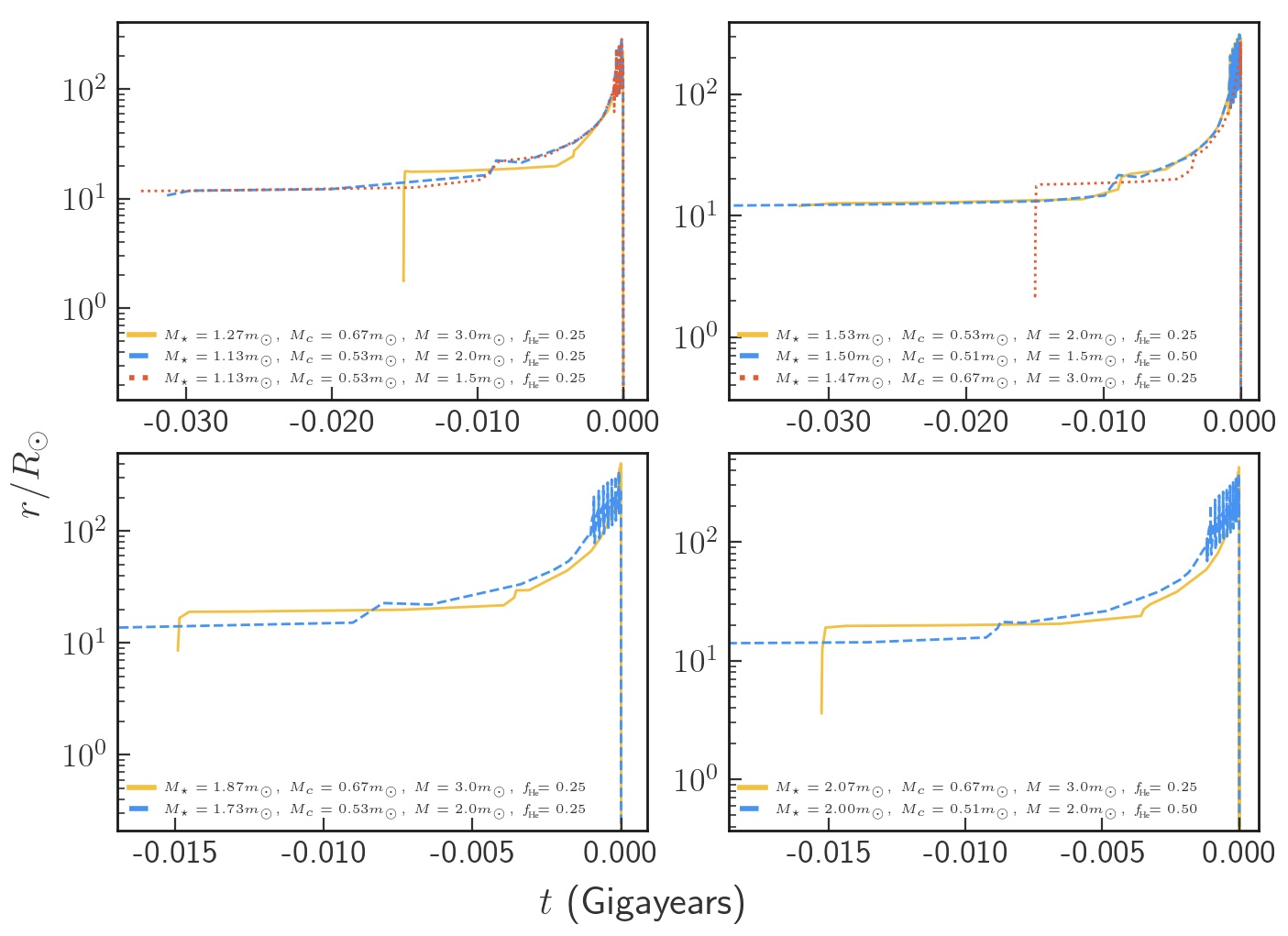}
    \centering
    \caption{Radius vs. time of giants with { (approximately)} the same {  mass after the stripping, $M_*$.} The time coordinate has been shifted so that they all become white dwarfs at the same time which is set as t=0. For stripped stars we show only the evolution after the stripping moment that is characterized by an initial very fast rise in the size. }
    \label{fig:rad_time_totalmass}
\end{figure*}

\begin{figure}
\includegraphics[width=.48\textwidth]{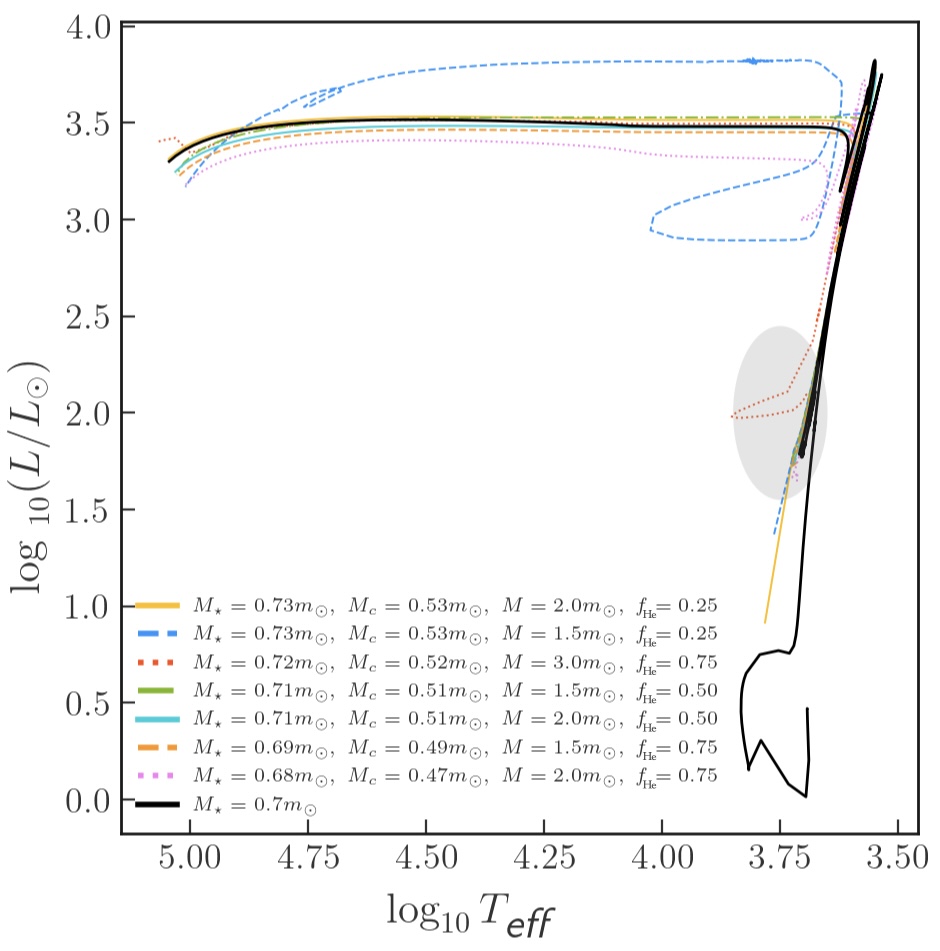}
    \centering
    \caption{HR diagram tracks for stars of the same mass after  stripping:  $M_\star=0.7\pm 0.03 m_\odot$   and an unstripped preZAMS $m_\odot$ star (black, thick line). Those include stars of different initial masses, $M$,  that were stripped at different $f_{\rm He}$ values.  The location of the different giants at the time of stripping is marked by the shaded area.
    }
    \label{fig:hr_070}
\end{figure}

As shown in Fig. \ref{fig:new}   stars stripped at later stages ($f_{_{\rm He}} =0.25$) die earlier than their progenitor stars, since they are unable to sustain nuclear burning. However, stars stripped earlier (at $f_{_{\rm He}}=0.75$ ) die later than their progenitor star. They still have plenty of helium to burn but their total mass has been reduced, so they behave like lower mass giants that burn He slightly slower and live longer. The transition between both situations is  at $f_{_{\rm He}} \approx 0.50$ for which some stripped cases live longer than the progenitor stars and some stripped cases have their lives shortened. 


The differences in the lifetimes of stripped stars when compared to their progenitor star, be it an earlier death or a prolonged life, are of the order of $\sim 1-5\times10^6$ years {{ (see Fig.~\ref{fig:new})}}. Compared to the duration of the giant phase ($\sim {\rm a ~ few} \times 10^8 ~{\rm yr} $  for a $0.1Z_\odot$,   $3 m_\odot$ star and $\sim {\rm a ~ few} \times 10^9 ~{\rm yr} $  for a $1m_\odot$ star), it constitutes a brief and insignificant period, especially for lower mass stars.

In Fig.~\ref{fig:rad_time_totalmass} we show the radius vs. time (shifted so that all stars become white dwarfs at the same time) for four different stripped stars and unstripped stars with the same total mass. Again, the radii are all similar, regardless of the progenitor star. The reason for this similar behavior is quite simple: in the range we consider the cores' masses depend very weakly on the total mass of the unstripped stars. Hence, after stripping, resulting stars that have the same total mass are essentially similar. 
A slight difference in the overall behavior is the appearance of late-stage thermal pulsations for all stars exceeding $1m_\odot$. 

Fig.~\ref{fig:hr_070} depicts the HR diagram  tracks of stars with total mass $M_\star = (0.7\pm0.03)m_\odot$ after stripping, compared to the track of a unstripped pre-zero age MS $m_\odot$ giant star (black thick line). The luminosity of the stripped stars is higher by a factor of $\sim 2$ (as compared to an unstripped star with the same total mass). This is because the stripped stars' cores are  more massive.  
Some tracks in Fig.~\ref{fig:hr_070},  show loops in the planetary nebula  phase. This are late thermal pulses, consisting of  a helium shell flash after the AGB phase that causes a rapid looping evolution between the AGB and planetary nebula phase. It should be regarded as an artifact of the parameters of the simulation and not as a characteristic feature of the stripped star, since the results at these stages are very sensitive to input physics.

\subsection{Successive PTDEs} 

An interesting possibility that might be common is a repeated sequence of partial TDEs. The remnant's trajectory  depends on the momentum loss during the disruption \citep[see ][for MS stars]{ryuTidalDisruptionsMainsequence2020b}.
{ The momentum loss  is of order 
$\Delta M v_{\rm esc} \approx  \sqrt{G M_*/r} $, where $v_{\rm esc}$ is the escape velocity from the disrupted boundary layer, $r$ (see Eq. \ref{eq:mr}). 
This is much smaller than  the remnant's orbital  momentum  $M_* \sqrt{(G M_{_{\rm BH}}/r_{\rm p})}$.  Hence we expect that }
the remnant's trajectory won't change significantly during the partial disruption  it will return to the vicinity of the SMBH with the same $r_{\rm p}$. A typical progenitor plunges towards the SMBH from the radius of influence, $R_{\rm h}$. Its remnant will return within $ 2 \pi R_{\rm h}^{3/2} /(G M_{_{\rm BH}})^{1/2}   \approx 2.5 \times 10^5~{\rm yr}$ where   we have used typical parameters of $R_{\rm h}=2$pc and $M_{_{\rm BH}}=10^6 m_\odot$. If the remnant has relaxed at this stage to a giant\footnote{Note that this orbital time is not much larger than the thermal time scale.} it will undergo a second PTDE at the same pericenter. As the envelope is smaller now, less mass will be torn apart in this second partial disruption (see Fig.~\ref{fig:td_bhmass1}).  The remnant will continue to a third disruption and so on.   We can expect a series of PTDEs.  As the envelope becomes  more dilute less mass is torn apart and the process converges to a light ({ $\sim0.7 m_\odot$ for the example shown in Fig.~\ref{fig:succesive}}) giant. Namely, the series of successive PTDEs don't strip the giant completely, and the remnant continues to evolve as a light giant. 

Fig.~\ref{fig:succesive} describes a Kippenhahn diagram for a series of successive disruptions of a $2 m_\odot$ giant by a $10^6 m_\odot$ SMBH. The PTDEs take place at the same pericenter. After each PTDE, the remnant returns to a giant structure. Then, it is disrupted again. The mass that is torn apart is estimated using Eq.~\ref{eq:mr} (see Fig.~\ref{fig:td_bhmass1}).  Given the series of disruptions, a significant fraction of the total mass is torn apart even though the pericenter is relatively large ($\sim 130 r_{\rm g}$).     

\begin{figure}
\includegraphics[width=0.48\textwidth]{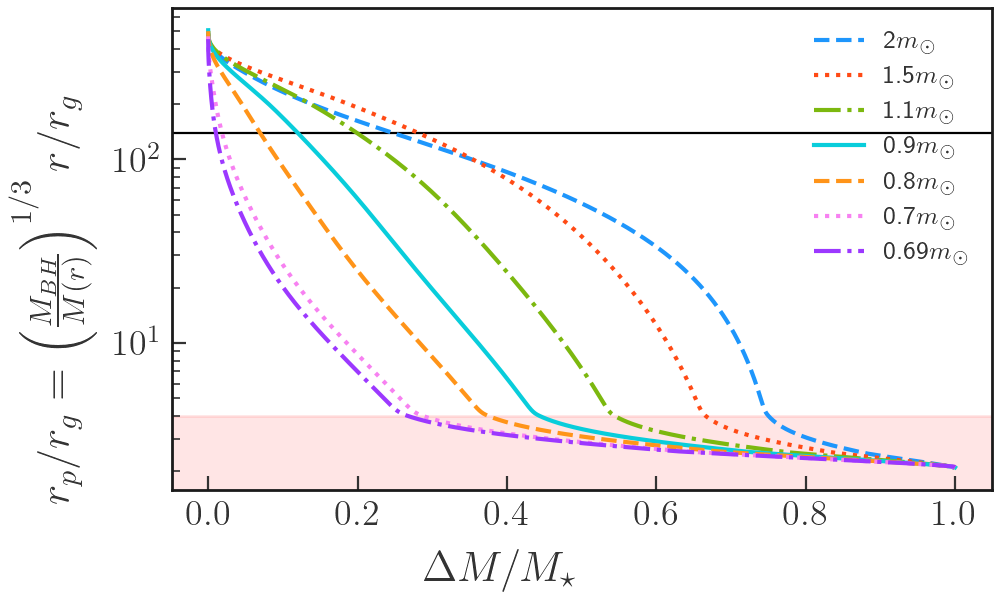}
\centering
    \caption{Stripped mass fraction, $\Delta M/M$, vs. $r_{\rm p} $ (in units of $r_{\rm g}$) for different giants that arise in successive PTDEs of a $2 m_\odot$ giant by a $10^6 m_\odot$ SMBH. The disruptions occur at $130 r_{\rm g}$, marked by a horizontal line. The red region marks $4 \rg$, which is the minimal pericenter distance for an orbit with zero orbital energy around a Schwarzchild black hole. The masses torn are $\Delta M = 0.5, 0.4  0.2, 0.1, 0.1  $ and $ 0.01 m_\odot$. The remnant masses are 1.5, 1.1, 0.9, 0.8, 0.7 and $0.69 m_\odot$. }
    \label{fig:td_bhmass1}
\end{figure}

\begin{figure}
\includegraphics[width=0.48\textwidth]{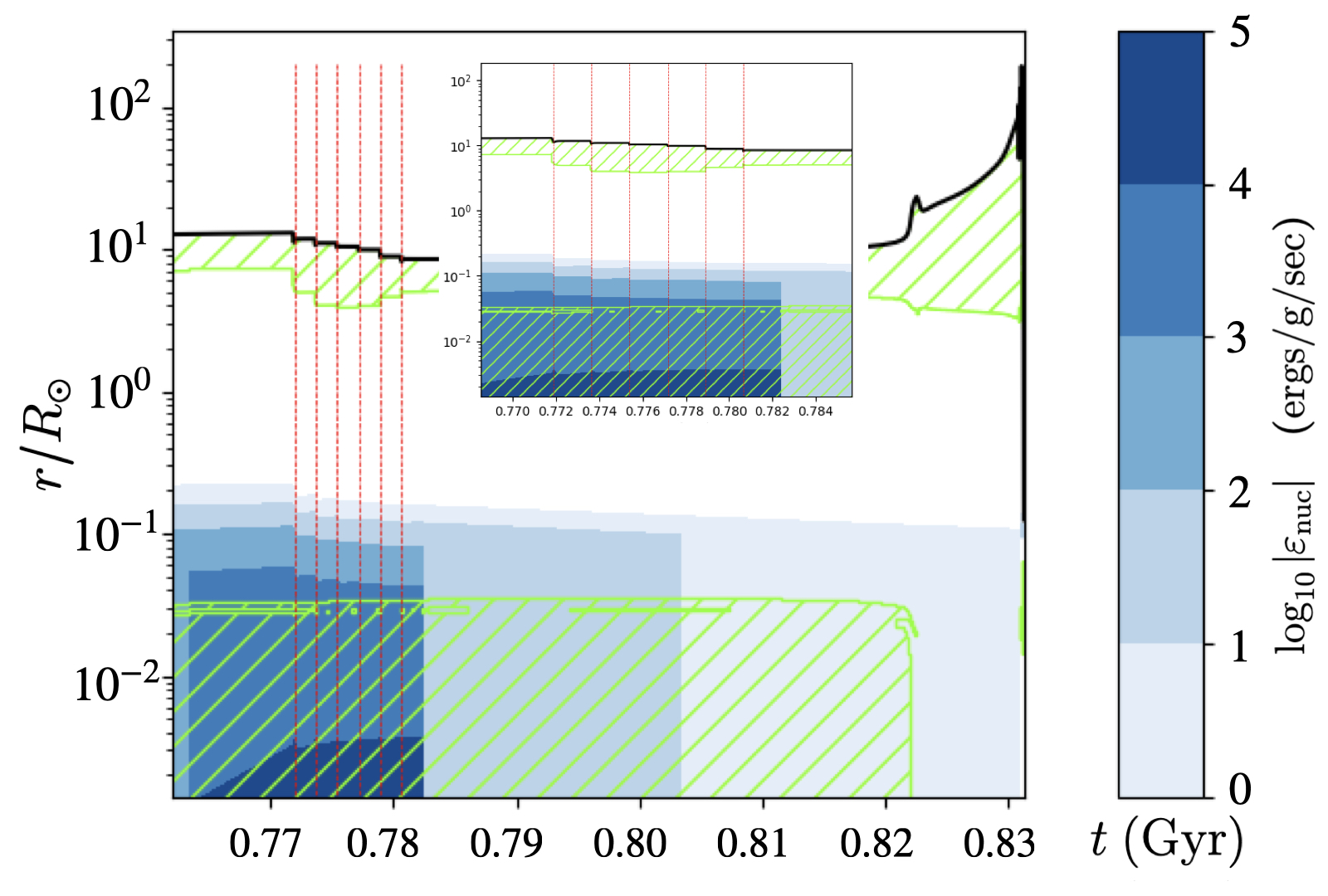}
    \centering
    \caption{A Kippenhahn diagram of a $2 m_\odot$ giant that undergoes successive PTDE (marked by dashed rad lines) at the same pericenter of $\approx 130 r_{\rm g}$. The time between disruptions is $2 \times 10^6 ~{\rm yr}$. The remnant returns each time to a giant stage (after $\sim 6 \times 10^4 ~{\rm yr}$ and then the disrupted mass is calculated using Eq.~\ref{eq:mr}
    (see also Fig.~\ref{fig:td_bhmass1}). Remnant masses are 1.5, 1.1, 0.9, 0.8, 0.7 and $0.69 m_\odot$. The insert describes the details around the stripping moments. }
    \label{fig:succesive}
\end{figure}

The overall evolution of the disrupted star is not very different from the evolution of the progenitor or from the evolution of a giant from which a larger mass was torn apart in a single event (see Fig.~\ref{fig:succesive} and middle panel of Fig.~\ref{fig:kipp_stripped}).  This can be understood in terms of earlier results. We have seen that the PTDE doesn't change significantly the remnant giant's lifetime as compared to the progenitor's lifetime. The lifetime is determined by processes taking place around the core, which are not influenced significantly by the partial disruption. 

\section{Conclusions}
\label{sec:conclusions}

We highlight the importance of partial tidal disruption events of giants within the broader category of tidal disruption events.  In particular,  most (all for $M_{\rm BH} \gtrsim 3 \times 10^5 m_\odot$) TDEs of giants strip only a small fraction of the giant's envelope. 
The deep potential well of the dense giants' cores enables them to retain a significant portion of their envelopes even in a nearby passage of a giant black hole. This motivates exploration of the fate of the remnants. 

We  have considered giants on the horizontal branch that are burning He at their cores.  For more massive stars ($M \gtrsim 1.5 m_\odot$) this is the longest 
among the different giant stages (see Fig.~\ref{fig:radius_time_all}). 
For less massive stars the subgiant phase is longer but the stellar radius is smaller at that phase. 
AGB stars and red giants  are larger by up to factor of $\sim 10$ than these giants and  the cross section for AGB stars and red giants for a  TDE  is larger by up to a factor of $\sim 10$ (depending on the question whether the loss-cone is full or empty). However, the duration of these phases is shorter by about a factor of $\sim 100$ than the duration of the He core burning phase (see Fig.~\ref{fig:radius_time_all}). Overall TDEs at these stages are less likely by a factor of $\sim 10$. Still it is interesting to explore in the future the fate of TDEs at this stage. It is important, however, when considering ``typical" giant TDEs to consider stellar parameters of the He core burning phase and not the extreme values of AGBs or red giants. 

We evolve realistic giant stars with the {\texttt {MESA}} stellar evolution code. We strip part of the envelope, treating the stripping process as an artificial strong wind, assuming spherical symmetry. Because of {\texttt {MESA}}'s limitations, the stripping period is much longer than the actual duration of the TDE. As a result, we obtain the structure of the relaxed remnant  after a thermal time scale. We then follow the rest of the evolution until the collapse to a WD. 

We expect that during this brief period of about $10^5 ~{\rm yr}$ that was not accurately simulated in {\texttt {MESA}} the star would move to the left and down in the HR diagram. The motion down is because the luminosity of the stripped star is slightly lower than the luminosity of the progenitor (before stripping). As the envelope is stripped, the outer radius of the star decreases, and its effective temperature increases.   Depending on the overall rate of TDEs we expect at most one remnant in such a stage in a galaxy. Once the star expands and relaxes to a giant it moves horizontally back to the regular giant branch.

We summarize our main findings:
\begin{itemize}
\item The stripping of a significant mass of the giant takes place only if $r_{\rm p}$, the pericenter of the giant's orbit, is at least a factor of 2 smaller than the characteristic tidal radius, $R_{\rm t}$, in which tidal stripping begins. Estimates of the return time of the bound material or the energy of the unbound material should take this factor into account. 

\item The stripped stars, despite a reduction in mass, expand and  maintain a similar radius to their progenitors. These remnants exhibit slightly more massive and denser cores relative to the total stellar mass alongside a more tenuous envelope.

\item There is a small ($\sim { \rm a~ few } ~\times 10^6 {\rm yr}$) difference in the lifetime of stripped stars compared to regular giants of the same mass depending on the time of stripping. The stripped stars die earlier (later) if stripped at the final (early) stages of their giant phase. 

\item The stripped stars have a higher (by a factor of $\sim 2$) luminosity,  as compared to a regular giant with the same total mass. This is due to their slightly larger cores. This may not  be sufficient to distinguish these remnants from regular giant stars just by their position on the HR diagram. 

\item Low-mass stars, $< 0.9m_\odot$, cannot reach the giant stage within the age of the Universe. The presence of such stars  indicates their origin as remnants of higher-mass stars that have undergone stripping. Asteroseismology measurements can identify such stars.  Indeed  low mass He burning giants have been identified in the Galaxy \citep{slimmerrg2022,Matteuzzi2023}. However, other potential stripping mechanisms, in particular binary tidal interaction,  can lead to similar remnants. The existence of such stripped giants in the field   \citep{slimmerrg2022,Matteuzzi2023} suggests that these alternative processes take place. To search for remnants of PTDEs we have to look at the central regions of our Galaxy as we may expect about a hundred giant  PTDE remnants in the Galactic center. It may be too difficult to identify such remnants in distant galaxies. 

\item An interesting possibility  is a repeated sequence of PTDEs. If the disruption doesn't change significantly the remnant's trajectory it  will return to the vicinity of the SMBH. As the remnant is relaxed at this stage to a giant configuration it will undergo a second PTDE.  As the disrupted star has a lower mass, less mass will be torn apart (see Figs.~\ref{fig:td_bhmass} and~\ref{fig:td_bhmass1}). Overall, we can expect a series of PTDEs in which less and less mass is stripped until the PTDEs become ineffective and the star ends as a light, $ \sim 0.6-0.7 m_\odot$ giant. Remarkably, the successive PTDEs don't significantly change the lifetime of the giant before it collapses to a WD. This can be understood as the PTDEs not influencing the core that determines this.

\end{itemize}

Before concluding we turn to  the rate of such tidal events. On the one hand giants are rare compared to MS stars with a ratio of $\sim 1:10 $  arising from the ratios of lifetimes on the respective branches. On the other hand the larger giant radii  make for larger tidal radii than those of MS stars. As a result, the loss cone is bigger. This increases the rates of per giant partial TDE rate approximately by a factor of $\sim R_{_{\rm Giant}}/R_{_{\rm MS}} \sim 10 $ in the full loss cone regime and $\sim \ln(R_{_{\rm Giant}}/R_{_{\rm MS}})$ in the empty loss cone regime.  
Furthermore, as first discussed in \citet{syerTidalDisruptionRates1999}, when a star becomes a giant its radius increases and hence stars can ``grow into the loss cone",  increasing the rate of disruptions.
Everything combined, the ratio between giant stars and main sequence stars TDE rates is approximately $\dot{N}_{\text{Giant}}/\dot{N}_{\text{MS}} \sim 10\%$  \citep{magorrianRatesTidalDisruption1999, macleodTidalDisruptionGiant2012}, a non negligible fraction of all TDEs. Assuming a conservative rate of a TDE per $10^4-10^5$ yr in our Galaxy and given a lifetime of a few $\times 10^7$ yr we expect a few dozens to a few hundred  stripped giant remnants in the Milky Way center. It is an interesting observational challenge to identify them.

Finally, we note that while we were considering TDEs by SMBHs at galactic centers, our results can  also be relevant in the context of  IMBHs within globular clusters in which abundance of remnants of PTDEs of giants around the core could indicate the presence of an IMBH. Here one has to carefully consider also the possibility  that stars in globular clusters may also undergo stripping due to interactions with other compact objects or heavier stars within the cluster.

We thank Sivan Ginzburg, Julian Krolik and Taeho Ryu for  constructive comments and 
Rob Farmer,  Selma de Mink and Mathieu Renzo, for helpful advice. 
The research was supported by an ERC grant, MultiJets, and by the Simons Collaboration on Extreme Electrodynamics of Compact Sources (SCEECS).

\section*{Appendix }
We briefly outline our implementation of \texttt{MESA}.

The \texttt {MESA} EOS is a blend of the OPAL \citep{Rogers2002}, SCVH
\citep{Saumon1995}, FreeEOS \citep{Irwin2004}, HELM \citep{Timmes2000}, PC \citep{Potekhin2010}, and Skye \citep{Jermyn2021} EOSes.

Radiative opacities are primarily from OPAL \citep{Iglesias1993, Iglesias1996}, with low-temperature data from \citet{Ferguson2005} and the high-temperature, Compton-scattering dominated regime by \citet{Poutanen2017}.  Electron conduction opacities are from \citet{Cassisi2007} and \citet{Blouin2020}.

Nuclear reaction rates are from JINA REACLIB \citep{Cyburt2010}, NACRE \citep{Angulo1999} and additional tabulated weak reaction rates \citep{Fuller1985, Oda1994, Langanke2000}.  Screening is included via the prescription of \citet{Chugunov2007}. Thermal neutrino loss rates are from \citet{Itoh1996}.

Convection is treated with the standard MLT of \citet{CoxGiuli}. We use the Schwarzschild criterion for convective stability and the predictive mixing scheme, following \citet{Ostrowski_2020} to avoid core splitting. We do not implement overshoot.

The Helium flash and thermally pulsing AGB (TPAGB) stages of evolution, if reached, are especially challenging. To reach numerical convergence, during those stages, we use \texttt{eps\_grav} as the \texttt{energy\_eqn\_option}, set to true
\texttt{convergence\_\linebreak ignore\_equL\_residuals} and enable MLT++ with
\texttt{okay\_to\_reduce\_gradT\_excess}.

We use \citet{reimersCircumstellarAbsorptionLines1975} for mass loss during the red giant phase, with a scaling factor $\eta_R=0.477$ following \citet{mcdonaldMassLossRed2015}. For the mass loss in the AGB phase we use \citet{bloeckerStellarEvolutionLow1995} with a scaling factor $\eta_B=0.1$.

To explore numerical convergence, we performed a series of resolution tests. We changed both spatial resolution with \texttt{mesh\_delta\_coeff} and temporal resolution with \texttt{time\_delta\_coeff} with values from $0.4$ to $1.8$ in steps of $0.2$. Some of the final tests are plotted in Fig. ~\ref{fig:tempres}.
We chose to work with the parameters \texttt{mesh\_delta\_coeff} = 0.6 and \texttt{time\_delta\_coeff} = 1.

\begin{figure}[h]
\includegraphics[width=.45\textwidth]{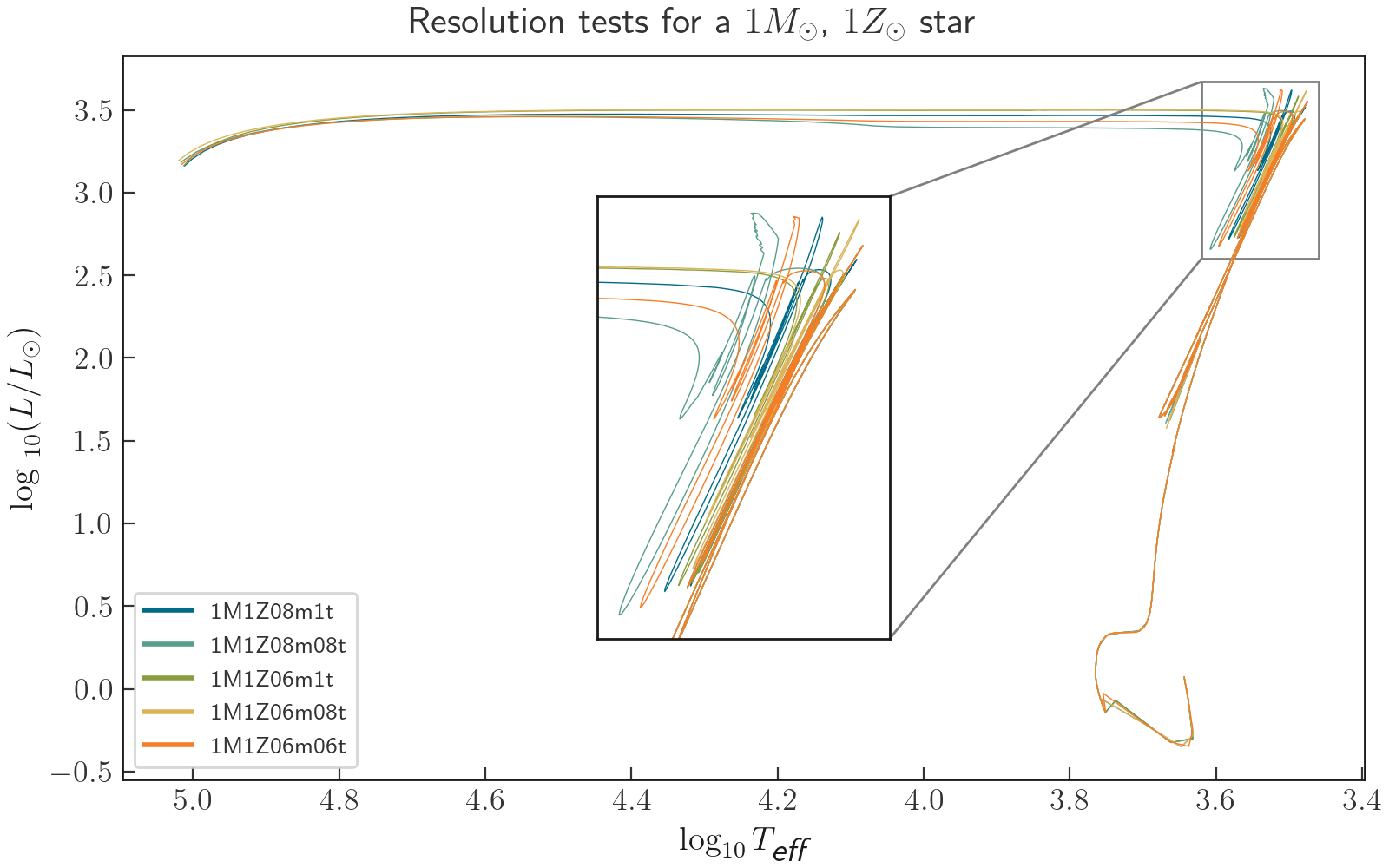}
    \centering
    \caption{HR diagram track of a $M=1m_\odot$, $1Z_\odot$ star from pre zero-age MS to WD for different temporal and spatial resolution values. The chosen final values correspond to the green line, mesh=0.6 and time=1.}
    \label{fig:tempres}
\end{figure}

The stripping process is performed using the {\texttt {MESA}} flags \texttt{relax\_initial\_mass\_to\_remove\_H\_envelope} and \texttt{extra\_mass\_retained\_by\_remove\_H\_env}, equivalent to relaxing the mass to \texttt{new\_mass = He\_core\_mass + extra\_mass\_retained\_by\_remove\_H\_env}. 

We set \texttt{lg\_max\_abs\_mdot = 0}.

The inlists used can be downloaded from \href{https://zenodo.org/records/10156193}{https://doi.org/10.5281/zenodo.10156193}

\bibliographystyle{aasjournal}

\end{document}